# Probing the Charge Separation Process on In$_2$S$_3$/Pt-TiO$_2$ Nanocomposites for Boosted Visible-light Photocatalytic Hydrogen Production

Fenglong Wang [a, b], Zuanming Jin [b, c], Yijiao Jiang [d,*], Ellen H. G. Backus [b], Mischa Bonn [b], Shi Nee Lou [a], Dmitry Turchinovich [b] and Rose Amal [a, *]

[a] School of Chemical Engineering, UNSW Australia, Sydney, NSW 2052, Australia.

[b] Max Planck Institute for Polymer Research, Ackermannweg 10, 55128 Mainz, Germany.

[c] Department of Physics, Shanghai University, 99 Shangda Road, Shanghai 200444, China.

[d] Department of Engineering, Macquarie University, Sydney, NSW 2109, Australia.

AUTHOR INFORMATION

**Corresponding Author**

Tel: +612-9385-4361, Email: r.amal@unsw.edu.au; yijiao.jiang@mq.edu.au

**Abstract**

A simple refluxing wet-chemical approach is employed for fabricating $In_2S_3$/Pt-$TiO_2$ heterogeneous catalysts for hydrogen generation under visible light irradiation. When the mass ratio between Pt-$TiO_2$ and cubic-phased $In_2S_3$ (denoted as $In_2S_3$/Pt-$TiO_2$) is two, the composite catalyst shows the highest hydrogen production, which exhibits an 82-fold enhancement over *in-situ* deposited Pt-$In_2S_3$. UV-vis diffuse reflectance and valence band X-ray photoelectron spectra elucidate that the conduction band of $In_2S_3$ is 0.3 eV more negative compared to that of $TiO_2$, favoring charge separation in the nanocomposites. Photoelectrochemical transient photo-current measurements and optical pump - terahertz probe spectroscopic studies further corroborate the charge separation in $In_2S_3$/Pt-$TiO_2$. The migration of photo-induced electrons from the $In_2S_3$ conduction band to the $TiO_2$ conduction band and subsequently into the Pt nanoparticles is found to occur within 5 picoseconds. Based on the experimental evidence, a charge separation process is proposed which accounts for the enhanced activity exhibited by the $In_2S_3$/Pt-$TiO_2$ composite catalysts.

# 1. Introduction

The photocatalytic solar-to-hydrogen conversion process is regarded as a potential solution to mitigate our dependence on non-renewable fossil fuels.[1-5] Titanium dioxide ($TiO_2$) is one of the most studied photocatalysts because of its ready availability, high photo-stability and low toxicity.[6-9] Despite these attributes, the large band gap of $TiO_2$ (~3.2 eV) has impeded its widespread applications as it limits the effective use of solar radiation to 3-4 %, since only the UV spectral component can be utilized.[10] It is highly desirable to extend the absorption of $TiO_2$ materials into the visible light region to make more effective use of the available solar spectrum. A promising way to extend absorption into the visible light region is to sensitize $TiO_2$ with narrow band-gap semiconductors that have a suitable band alignment with $TiO_2$.[11] For this purpose, metal sulfides have been extensively investigated. Among them, cadmium sulfide (CdS), with a band gap of *ca.* 2.3 eV and a more negative conduction band compared to that of $TiO_2$, is one of the most studied candidates for sensitizing $TiO_2$ for hydrogen generation.[12] However, cadmium is a toxic heavy metal, so there is a large incentive to explore less toxic metal sulfides to sensitize $TiO_2$. Recently, due to its lower toxicity, indium sulfide ($In_2S_3$) has attracted considerable interest for hydrogen generation under visible light irradiation.[13] $In_2S_3$-based photocatalysts, however, are known to be susceptible to photocorrosion but can operate efficiently in the presence of sacrificial electron donors. [14, 15] Chai *et al.* have reported composites of $In_2S_3$ and Pt-$TiO_2$ prepared by a ball milling method for hydrogen generation.[16] They found that after coupling with Pt-deposited Degussa P25 $TiO_2$, the $In_2S_3$/Pt-$TiO_2$ nanocomposites showed enhanced hydrogen generation compared to Pt-deposited $In_2S_3$. These previous research indicated that $In_2S_3$ can act as an efficient sensitizer for $TiO_2$ catalysts to capture wider solar energy spectra.

However, an energy-effective chemical approach for fabrication of such a composite catalyst is not yet to be reported.

In addition, unravelling the dynamic behaviour of the photo-excited electrons is of significant importance in order to advance the understanding of the electronic properties of photocatalysts to develop efficient hybrid systems in the future. Kamat et al. found that the photo-excited electrons on CdSe could transfer to graphene oxide nano sheets within several nanosecond (ns) by observing its photoluminescence quenching.[17] They also discovered that the electron transfer rate from $Ag_8$ cluster to $MV^{2+}$ was $2.74\times10^{10}$ $s^{-1}$, confirming the ability of such metal clusters to participate into photocatalytic reduction reactions.[18] By using transient absorption spectroscopy their group further found that in the CdSe-squaraine dye-$TiO_2$ nanocomposites, both the electron migrations from CdSe to squaraine dye and from squaraine dye to $TiO_2$ nanoparticles took place on picosecond (ps) time scale while Lian's group found that the photo-excited electron transfer from CdS nanorods to Pt nanoparticles occurred at ~3.4 ps.[19-21] Recently, optical pump - terahertz (THz) probe (OPTP) spectroscopy emerges as a powerful tool to probe the dynamics and trapping/capturing effect of the photo-excited charge carriers in photocatalysts for solar fuels production, due to its high sensitivity to the mobile electrons.[22] Meng et al. observed the electron transfer from hematite to reduced graphene oxide occurring in several ps with the time-domain terahertz spectroscopy.[23]

Herein, we report the fabrication of $In_2S_3$/Pt-$TiO_2$ nanocomposites for visible-light photocatalytic hydrogen generation through an energy-efficient refluxing method. It is found that the $In_2S_3$/Pt-$TiO_2$ composite, with a Pt-$TiO_2$:$In_2S_3$ mass ratio of two shows the highest hydrogen yield - over 80 times higher than that of the *in-situ*-deposited Pt-$In_2S_3$. UV-vis diffuse reflectance spectra combined with valence band X-ray photoelectron spectroscopy (XPS) spectra are used to reveal a favorable conduction band alignment between $In_2S_3$ and $TiO_2$. Photoelectrochemical (PEC) transient photocurrent analysis is used to highlight an efficient charge separation process at the heterogeneous $In_2S_3$/Pt-$TiO_2$ interface. Additionally, OPTP spectroscopic studies are employed to show that photo-generated electrons in the $In_2S_3$ conduction band can be injected into the $TiO_2$ conduction band and then into the Pt nanoparticles. The charge transfer is found to be as fast as only 5 ps.

## 2. Experimental

**2.1 Preparation of Pt-$TiO_2$ catalysts** Bare $TiO_2$ nanoparticles were synthesized using an evaporation-induced self-assembly method as reported previously.[24] Briefly, 10 ml of titanium isopropoxide (TTIP) was added dropwise to 250 ml of Milli-Q water under vigorous stirring, with the suspension then sonicated for 30 min before aging for two days. Next, the water was evaporated in air at 100 °C to obtain a white powder which was calcined at 500 °C for 1 h to give crystallized $TiO_2$ particles. A chemical reduction method was utilized to prepare the Pt-decorated $TiO_2$ composites with a metal content of 1.0 wt. %. Firstly, a given amount of chloroplatinic acid hydrate was dissolved in 500 ml of Milli-Q water under stirring, followed by the addition of 5 g of the synthesized $TiO_2$ particles. The suspension was stirred for a further 1.5 h in an ice bath to achieve sufficient contact between the metal precursor and $TiO_2$ particles. Subsequently, a surplus of ice-cold $NaBH_4$ solution (0.1 M) was quickly injected into the suspension under vigorous stirring. The suspension was stirred for another 3 h and the products were washed with ethanol and water. The obtained solids were dried at 110 °C overnight. No further thermal treatment was applied to the resulting grey powder, that is denoted as Pt-$TiO_2$.

**2.2 Preparation of $In_2S_3$/Pt-$TiO_2$ nanocomposites** In a typical process, 1.0 g of the prepared Pt-$TiO_2$ catalysts was dispersed in 100 ml of Milli-Q water, followed by the addition of a given amount of indium nitrate ($In(NO_3)_3$) and thioacetamide (TAA). The suspension was sonicated for 10 min and stirred for 30 min. Subsequently, the suspension was refluxed at 95 °C for 5 h under stirring, with the product then washed and dried for further use. The molar ratio of In(NO$_3$)$_3$ and TAA was 1:2.5 and the nominal weight ratios of In$_2$S$_3$ and Pt-TiO$_2$ were set at 1:10, 1:5, 1:2 and 3:4. The In$_2$S$_3$/TiO$_2$ nanocomposite control was prepared using the same method as for the In$_2$S$_3$/Pt-TiO$_2$ nanocomposite whereby TiO$_2$ was present in place of Pt-TiO$_2$. Preparation of the neat In$_2$S$_3$ was conducted in the same maner but without Pt-TiO$_2$ addition.

**2.3 Characterization** X-ray diffraction (XRD) patterns were recorded on a Philips X'pert Multipurpose X-ray diffraction system using Cu K-alpha radiation in step mode between 20° and 80° with a step size of 0.026° and 50 s per step. The XRD operating conditions were 45 kV and 40 mA. High resolution transmission electron microscopy (HR-TEM) analysis was performed on a Phillips CM200 instrument with an accelerated voltage of 200 kV. UV-vis diffuse reflectance spectra (DRS) were obtained on a Shimadzu UV 3600 spectrophotometer. XPS were recorded on an ESCALAB250Xi spectrometer (Thermo Scientific, UK) applying a mono-chromated Al K-alpha source at 15.2 kV and 168 W.

**2.4 PEC measurements** The photo-electrodes were prepared by drop-casting. In a typical process, 6.0 mg of the catalyst was dispersed in 6 ml of ethanol by sonication and then the suspension was dropped onto fluorine-doped tin oxide glass slides (FTO, 2 cm×1 cm) through a layer-by-layer self-assembly method. Photocurrent measurements were undertaken at room temperature in a 0.1 M Na$_2$SO$_4$ aqueous solution as the electrolyte using an Autolab potentiostat (Model PGSTAT302N). A bias of 0.3 V was applied in a two-electrode PEC cell with Pt as the counter electrolyte and reference electrode and the catalyst-coated FTO glass as the working electrode. The electrolyte solution was purged with N$_2$ for 10 min prior to analysis with the purging continued during the photocurrent measurement to remove any dissolved O$_2$ in the cell. The photocurrent was measured using a 300 W Xenon lamp with a 420 nm cut-off filter. To minimize the heat effect of the electrolyte solution during illumination, a water jacket was placed between the Xe lamp and the PEC cell.

**2.5 OPTP** THz spectroscopic samples were fabricated via a drop-casting method using an aqueous suspension of each sample. Typically, 200 μl of the aqueous suspension with a concentration of 10 mg/ml was dropped on the surface of a fused silica substrate (1 cm×1 cm) and spread evenly. The water was then evaporated at 100 $^o$C using a hot plate to provide a reasonably uniform thin film. The set-up for OPTP measurement is detailed elsewhere.[25, 26] The OPTP measurement was conducted at room temperature in a $N_2$ atmosphere to minimize the oxidation effect and the interference from water vapour. The instrument was driven by a Ti:sapphire femtosecond amplifier, operating at a repetition rate of 1 kHz, and generating 120 fs pulses at a central wavelength of 800 nm. A portion of the laser output was doubled in a β-$BaB_2O_4$ (BBO) crystal. The resulting 400 nm pulse was used to excite the $In_2S_3$ as its energy is greater than the band gap of $In_2S_3$ but less than the band gap of $TiO_2$. A second portion of the laser was used to generate and detect the THz radiation by optical rectification and electro-optical sampling with ZnTe nonlinear crystals. Both optical pump and THz probe pulses were incident on the sample at normal incidence. The in-plane conductivity of the photo-excited sample was probed with the THz pulse. Time-resolved data was obtained by averaging 10 time-delay scans.

**2.6 Photocatalytic activity test** Visible-light photocatalytic hydrogen generation by the nanocomposites was evaluated using a top-irradiated reaction cell with a 300 W Xenon arc lamp equipped with an optical filter (λ > 420 nm). In a typical procedure, 300 mg of the prepared catalyst was dispersed in 200 ml of a $Na_2S$ (0.25 M) and $Na_2SO_3$ (0.25 M) solution by sonication. Argon was purged into the system for more than 30 min in order to completely remove air. The mixture was then irradiated by visible light under magnetic stirring and the temperature of the reaction system was maintained at *ca.* 20 $^o$C with cooling water flowing around the reaction vessel. The evolved gas was analyzed every 30 min using a gas chromatograph (GC, Shimadzu, 8A) equipped with a thermal conductivity detector (TCD). For the activity test using Pt/$In_2S_3$, 1.0 wt % of Pt was photo-deposited *in-situ* onto the $In_2S_3$ by injecting the Pt precursor into the suspension just prior to light irradiation. Repeating the experiments indicated errors associated with the activity test were *ca.* 5%.

## 3. Results and discussion

### 3.1 Texture studies

The crystallographic structure and phase texture of the composite materials were characterized by XRD, with the spectra displayed in Fig. 1. The XRD pattern of $In_2S_3$ exhibits cubic $\beta$ phase (JCPDS no. 65-0459).[27] No other phase was detected indicating the high purity of the prepared $In_2S_3$. The XRD pattern of Pt-$TiO_2$ shows the dominance of anatase $TiO_2$ with a trace amount of brookite $TiO_2$ as we reported previously.[24] It can be seen that with an increase in $In_2S_3$ content within the composites, the intensities of the diffraction peaks at 27.4 °, 33.2 °, 43.6 °, and 47.7°, indexed to the dominant (311), (400), (511) and (440) crystal planes of cubic $In_2S_3$, respectively, were increased. It is noted that no Pt diffraction peak was observed likely due to the fine particle size and good dispersion of the Pt nanoparticles on the $TiO_2$ surface.

The UV-vis diffuse reflectance spectra of the composite materials with various mass ratios were recorded to characterize their optical properties, as shown in Fig. 2. It can be seen that the bare $TiO_2$ nanoparticles only show efficient absorption in the UV light region with the absorption edge at 393 nm, corresponding to a band gap of 3.16 eV. After depositing the Pt, the particles displayed enhanced absorption in the visible light region, which is a typical optical property for Pt-$TiO_2$.[12] The bare $In_2S_3$ shows the strongest visible light absorption among all the samples and its band gap was estimated to be ~2.06 eV according to the equation of $E_g =1240/\lambda_g$, where $\lambda_g$ is the optical absorption edge of the semiconductor.[28] It is apparent that absorption by the $In_2S_3$/Pt-$TiO_2$ nanocomposites is gradually red-shifted with an increase in the $In_2S_3$ content. The shift is attributed to the visible-light response of $In_2S_3$. The results illustrate that the $In_2S_3$ presence can efficiently extend the absorption of $TiO_2$ nanoparticles into the visible light region.

HR-TEM images of Pt-$TiO_2$ and $In_2S_3$/Pt-$TiO_2$ nanocomposites are shown in Fig. 3. In the Pt-$TiO_2$ composites, Pt nanoparticles with a size of *ca.* 3 nm are well-dispersed on the $TiO_2$ surface (Fig. 3a). The inter-planar lattice spacing of 0.35 nm can be assigned to the (101) facet of anatase $TiO_2$. The *d*-spacing of 0.22 nm obtained on the Pt nanoparticle corresponds to the (111) facet of metallic Pt (Fig. 3a inset).[29] After coupling with $In_2S_3$, the Pt nanoparticle size remains constant (Fig.3b). The HR-TEM image also shows a phase junction between $In_2S_3$ and $TiO_2$, evidenced by the presence of the (101) facet of $TiO_2$ and (111) facet of $In_2S_3$ with the characteristic lattice spacings of 0.35 nm and 0.62 nm, respectively.[27] The TEM morphological information is in good agreement with the XRD patterns (Fig. 1) and confirms the successful preparation of $In_2S_3$/Pt-$TiO_2$ nanocomposites.

### 3.2 Photocatalytic hydrogen production performance

The photocatalytic activity of the as-synthesized materials was evaluated for hydrogen generation under visible light irradiation ($\lambda$ > 420 nm ≡ 2.95 eV) since $In_2S_3$ is visible-light-active as indicated in the UV-vis spectra (Fig. 2). A sacrificial $Na_2S$/$Na_2SO_3$ reagent mix was used to scavenge the photo-generated holes while the photo-induced electrons reduced protons into hydrogen. As shown in Fig. 4, the Pt-TiO$_2$ and In$_2$S$_3$/TiO$_2$ nanocomposites displayed no photocatalytic activity for hydrogen generation under visible light irradiation. The *in-situ* deposition of 1.0 wt. % of Pt on In$_2$S$_3$ provided some activity, resulting in a limited hydrogen yield of 7 µmol/g over 3 h. A similar phenomenon has previously been observed for CdS catalysts.[12] For the In$_2$S$_3$/Pt-TiO$_2$ composite materials, increasing the In$_2$S$_3$ and Pt-TiO$_2$ weight ratio from 1:10 to 1:2 increased the hydrogen production from 197 µmol/g to 573 µmol/g, which was over 80 times higher than hydrogen production by the *in-situ*-deposited Pt-In$_2$S$_3$. Increasing the In$_2$S$_3$ content beyond the 1:2 ratio led to a decrease in hydrogen production, which may originate from excess In$_2$S$_3$ impeding the sites needed for proton reduction.[11] Note that the In$_2$S$_3$ and TiO$_2$ composites exhibited no hydrogen production, illustrating the indispensable role of the Pt nanoparticles as co-catalysts to provide the active sites for proton reduction. As In$_2$S$_3$/Pt-TiO$_2$ with a weight ratio of 1:2 gave the best hydrogen yield under visible light irradiation, it was selected for further study of the electron transfer process at the heterojunction.

## 3.3 Charge separation studies

To understand the origin of the enhanced activity for hydrogen generation over the In$_2$S$_3$/Pt-TiO$_2$ nanocomposites, it is crucial to establish the band alignment between In$_2$S$_3$ and TiO$_2$. The valence band XPS spectra for In$_2$S$_3$ and TiO$_2$ are presented in Fig. 5a. From the spectra, it is estimated that the valence band maxima of In$_2$S$_3$ and TiO$_2$ lie at around 1.5 eV and 2.9 eV relative to the normal hydrogen electrode (NHE), respectively. It is known that the valence band of TiO$_2$ consists of O *2p* orbitals, of which the potential is about 3 eV *vs* NHE.[2, 30] It is reasonable that In$_2$S$_3$ shows a shallow valence band position since the level of S *3p* orbitals is always more negative than that of O *2p* orbitals.[2] The band gaps of In$_2$S$_3$ and TiO$_2$ are 2.06 eV and 3.16 eV, respectively, based on the UV-vis diffuse reflectance spectra (Fig. 2). Combined, these findings indicate that the conduction band minima of $In_2S_3$ and $TiO_2$ are located at -0.56 eV and -0.26 eV *vs* NHE, respectively. That is, the conduction band position of $In_2S_3$ is 0.3 eV more negative than that of $TiO_2$. A schematic diagram proposing the photocatalytic hydrogen generation process occurring on the $In_2S_3$/Pt-$TiO_2$ nanocomposites is shown in Fig. 5b. Upon irradiation ($\lambda > 420$ nm $\equiv$ 2.95 eV), visible light absorption by $In_2S_3$ gives rise to photo-generated electron-hole pairs, while the $TiO_2$ nanoparticles show no response due to its wide band gap (3.16 eV). The photo-excited electrons in the conduction band of $In_2S_3$ then transfer into the $TiO_2$ conduction band, driven by the potential drop at the interface. Finally, the electrons are captured by the Pt nanoparticles deposited on the $TiO_2$ surface. In an aqueous system, the Pt nanoparticles act as active sites for the proton reduction reaction, resulting in hydrogen formation. The holes remaining in the $In_2S_3$ valence band are involved in oxidation of the sacrificial reagent mix ($Na_2S_3$/$Na_2SO_3$).[11]

The scheme in Fig. 5b, detailing efficient charge separation within the $In_2S_3$/Pt-$TiO_2$ nanocomposites, is corroborated by PEC transient photo-current measurements in a two-electrode PEC cell. A fast and uniform photocurrent response was observed for the $In_2S_3$-containing samples for each on/off cycle under visible light irradiation. Fig. 6 clearly shows that, under visible light illumination, the photocurrent of the $In_2S_3$/Pt-$TiO_2$-coated electrode is strongly enhanced compared to that of the neat $In_2S_3$ electrode. A detailed evaluation of the photo-induced charge separation is shown in Fig. S1 which presents the photocurrent values on a per mg basis of the optically active $In_2S_3$ component. (See supporting information) Although Pt-$TiO_2$ shows an absorption tail in the visible light region (Fig. 2), no photocurrent is observed for this material, which is in line with the absence of photocatalytic activity.

Hence, the enhanced photocurrent exhibited by the $In_2S_3$/Pt-$TiO_2$ can be attributed to the improved separation of photo-induced charge carriers in the nanocomposites.

**3.4 Dynamics of photo-induced charge carriers**

In order to evaluate the ultrafast, picosecond-timescale relaxation and/or capture dynamics of the photo-induced charge carriers in the catalysts, OPTP spectroscopy was employed. OPTP spectroscopy allows for the quantitative investigation of sample conductivity, which is determined by the number of carriers and their mobility on very short (sub-picosecond) timescales following the optical injection of charge carriers. As a contact-free technique, OPTP is very well-suited for studying carrier dynamics in semiconducting nanostructures[31], and it has previously been successfully employed in photocatalytic studies.[22, 23] The time-dependent photoconductivity of the $TiO_2$, $In_2S_3$, $In_2S_3$/$TiO_2$ and $In_2S_3$/Pt-$TiO_2$ samples was recorded following their photo-excitation with an ultrashort excitation laser pulse of ~50 femtoseconds (fs) duration and a central wavelength of 400 nm. These transient photoconductivity measurements were performed by monitoring the change in amplitude of the peak of the THz probe pulse as a function of pump-probe time delay. Such a measurement is predominantly sensitive to the photo-induced change in the real part of the sample photoconductivity (which is in general complex-valued).[25, 32] Fig. 7a presents the normalized real photoconductivity dynamics for $In_2S_3$, $In_2S_3$/$TiO_2$ and $In_2S_3$/Pt-$TiO_2$. As shown in the inset of Fig. 7a, a near-instantaneous rise in the OPTP signal in $In_2S_3$ is observed, which indicates that free, mobile charge carriers (electrons and holes) are generated in $In_2S_3$ within ca. 1 ps after photoexcitation. In contrast, no OPTP signal is observed for $TiO_2$, which reveals that $TiO_2$ is not noticeably excited under the current experimental conditions owing to its larger band gap. After the initial rise, the real photoconductivity decays with a fast (few ps), a long-lived (tens of ps) and a very slow (> ns) component. The photoconductivity decay on a few-picosecond timescale is consistent with the picture of ultrafast electron trapping.[33] It is found that the OPTP signals for $In_2S_3$ and $In_2S_3$/$TiO_2$ are quite similar, which demonstrates that the addition of $TiO_2$ does not introduce any new trapping sites for the mobile electrons. Importantly, the carrier dynamics of $In_2S_3$/Pt-$TiO_2$ shows a significantly different trend, which can be attributed to the strong electron-capturing effect of the Pt nanoparticles.[34] The result agrees with the previous findings that Pt nanoparticles can efficiently trap the electrons from $TiO_2$ nanoparticles on ps timescales before they relax into the deep bulk trapping sites.[22, 34] By varying the Pt loading contents from 0 to 2 wt.% on $TiO_2$, Furube et al. observed enhanced electron migration from $TiO_2$ to Pt nanoparticles due to more Pt deposits on $TiO_2$ surface indicated by the more rapid decay of transient absorption spectra.[34] However, the shielding effect of larger Pt-loading was also obvious as the initial absorption was lower on the higher Pt supported samples. There would probably be competition between these two effects (electron trapping and shielding) that gives rise to the optimal Pt loading as observed in the photocatalytic hydrogen evolution studies.[35] In the current study, as the Pt nanoparticles are located on the $TiO_2$ surface, the above findings indicate that the photo-excited electrons in the conduction band of $In_2S_3$ flow to the $TiO_2$ conduction band and subsequently into the Pt nanoparticles on a ps time scale. This is consistent with the energetics of the system: the Fermi level of the Pt nanoparticles lies below the conduction band of $TiO_2$ which, in turn, possesses a positive potential of 0.3 eV relative to the conduction band of $In_2S_3$. Hence, the transfer of electrons from $In_2S_3$ to $TiO_2$, and then to Pt nanoparticles is energetically downhill.

To quantify the acceleration of the photoconductivity decay due to the electron transfer, bi-exponential fits, $\Delta E(t)=A_1*\exp(-t/\tau_1)+A_2*\exp(-t/\tau_2)+A_3$, are used to infer the decay time constants (circles) and amplitude ($A_1$ and $A_2$ bars) for the fast and slow processes, as shown in Fig. 7b and 7c, respectively. These reveal enhancements in both the amplitude (~18% vs 14%) and the rate (accelerated by a factor of 1.2) of the rapid decay in the $In_2S_3$/Pt-$TiO_2$ system; the slow decay is similarly accelerated by the presence of the Pt nanoparticles. While the fast decay corresponds to electron trapping, the slow decay with tens of ps time constant can also, at least in part, be attributed to interband recombination.[33, 36, 37] The above analysis shows that the photo-excited electrons in $In_2S_3$ can transfer to $TiO_2$ and further into Pt nanoparticles on a ~5 ps timescale in the ternary composites, which results in the greatly enhanced photocatalytic activity.

## 4. Conclusions

$In_2S_3$/Pt-$TiO_2$ nanocomposites with different weight ratios were successfully synthesized using a simple wet-chemical route. The $In_2S_3$ was found to be a good sensitizer for $TiO_2$, enabling hydrogen generation upon visible light illumination. At an optimal $In_2S_3$ to Pt-$TiO_2$ weight ratio of 1:2, the hydrogen yield was 82 times higher than that of the *in-situ* deposited Pt-$In_2S_3$. The band alignment scheme at the heterogeneous interface between $In_2S_3$ and $TiO_2$ was established using valence band XPS and UV-vis diffuse reflectance spectra to understand the origin of efficient charge transfer within the composite catalysts. The enhanced charge separation was further established by PEC transient photo-current analysis. OPTP spectroscopic studies on the composite photocatalysts revealed that the interfacial charge transfer occurred on a ~5 ps timescale. Overall, this work shows a novel, efficient route towards $TiO_2$-based composite materials for visible light photocatalysis, and provides fundamental insights into the photo-induced charge separation process.

**Acknowledgements**

Financial support by the ARC Discovery Early Career Researcher Award (DE120100329) and ARC Discovery Project (DP140102432) is gratefully acknowledged. The Mark Wainwright Analytical Centre, UNSW Australia is also acknowledged for providing analytical support to the project. FLW is grateful for the financial support from the Australia Nanotechnology Network (ANN) and German Academic Exchange Service (DAAD). EHGB and FLW thank the support provided by the ERC (Starting Grant No. 336679). ZJ and DT acknowledge the Max Planck Society and the European Commission (EU Career Integration Grant 334324 LIGHTER).

**Figure captions:**

**Fig. 1.** XRD patterns of Pt-TiO$_2$, In$_2$S$_3$, and In$_2$S$_3$/Pt-TiO$_2$ with different mass ratios.

**Fig. 2.** UV-vis diffuse reflectance spectra of TiO$_2$, Pt-TiO$_2$, In$_2$S$_3$, and In$_2$S$_3$/Pt-TiO$_2$ with different mass ratios.

**Fig. 3.** HR-TEM images of (a) Pt-TiO$_2$ and (b) In$_2$S$_3$/Pt-TiO$_2$ with mass ratio of 1:2. Inset of (a) shows the lattice fringe of the Pt (111) facet.

**Fig. 4.** Hydrogen production yield in aqueous solution by various In$_2$S$_3$/Pt-TiO$_2$ composite catalysts under visible light irradiation ($\lambda$ > 420 nm ≡ 2.95 eV), with Na$_2$S/Na$_2$SO$_3$ as the sacrificial reagent mix.

**Fig. 5.** (a) Valence band XPS spectra of In$_2$S$_3$ and TiO$_2$ and (b) proposed mechanism for photocatalytic hydrogen generation by the In$_2$S$_3$/Pt-TiO$_2$ composite under visible light irradiation and in the presence of a sacrificial reagent mix (Na$_2$S/Na$_2$SO$_3$).

**Fig. 6.** Photocurrent response of In$_2$S$_3$, Pt/TiO$_2$ and In$_2$S$_3$/Pt-TiO$_2$ nanocomposites under visible light irradiation ($\lambda$ > 420 nm). Every 80 s cycle comprises 20 s of irradiation on, followed by 60 s of dark current measurement.

**Fig. 7.** (a) Normalized THz photoconductivity signals for In$_2$S$_3$, In$_2$S$_3$/TiO$_2$, and In$_2$S$_3$/Pt-TiO$_2$ excited by a 400 nm pulse at room temperature. The inset shows the OPTP signals for In$_2$S$_3$ and TiO$_2$. (b) and (c) show the extracted amplitude (column) and decay times (circles) for the fast and slow decay components, respectively, for the three samples at a pump fluence of 500 μJ/cm$^2$.

**Figures:**

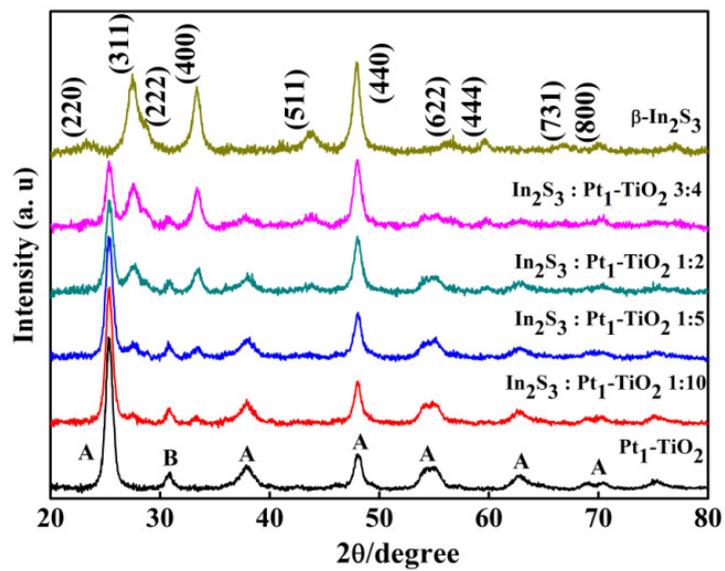

Fig. 1.

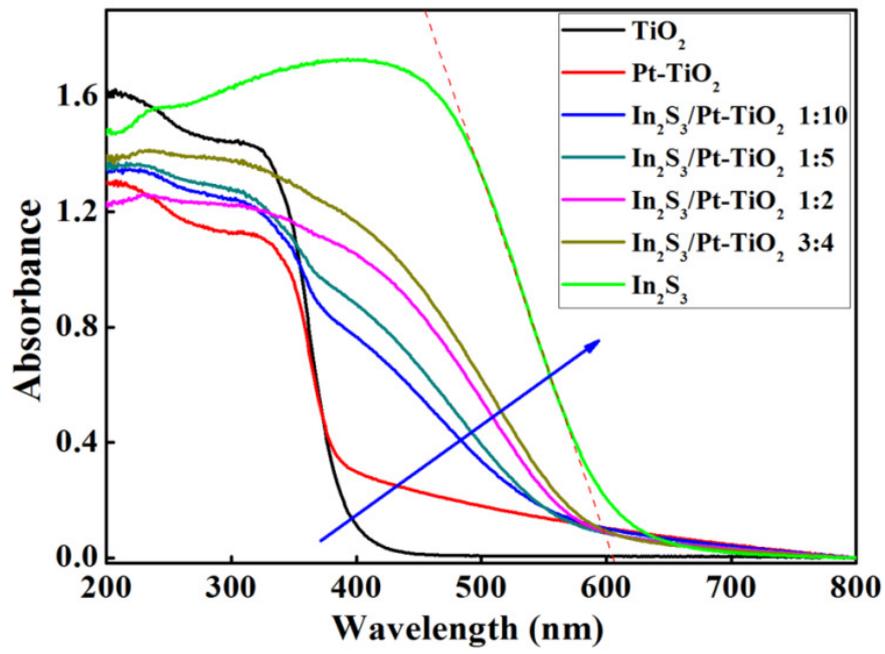

Fig. 2.

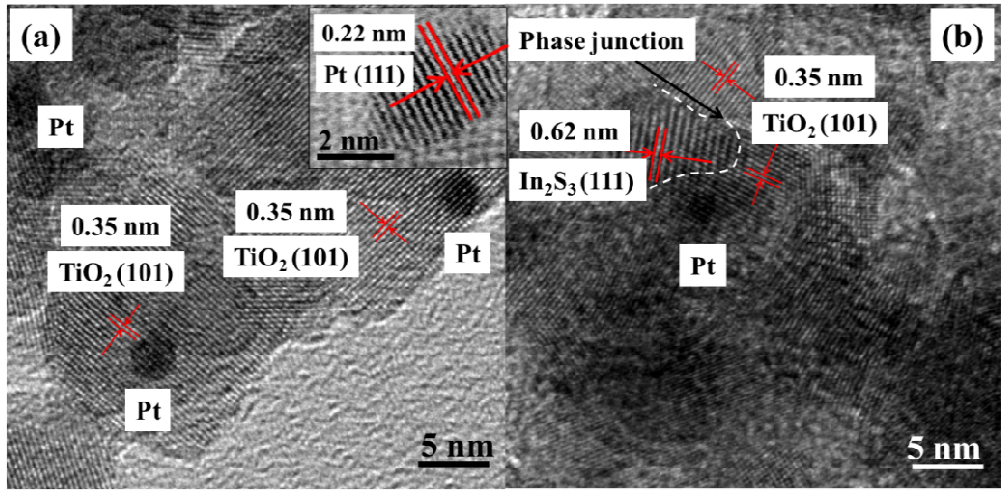

Fig. 3.

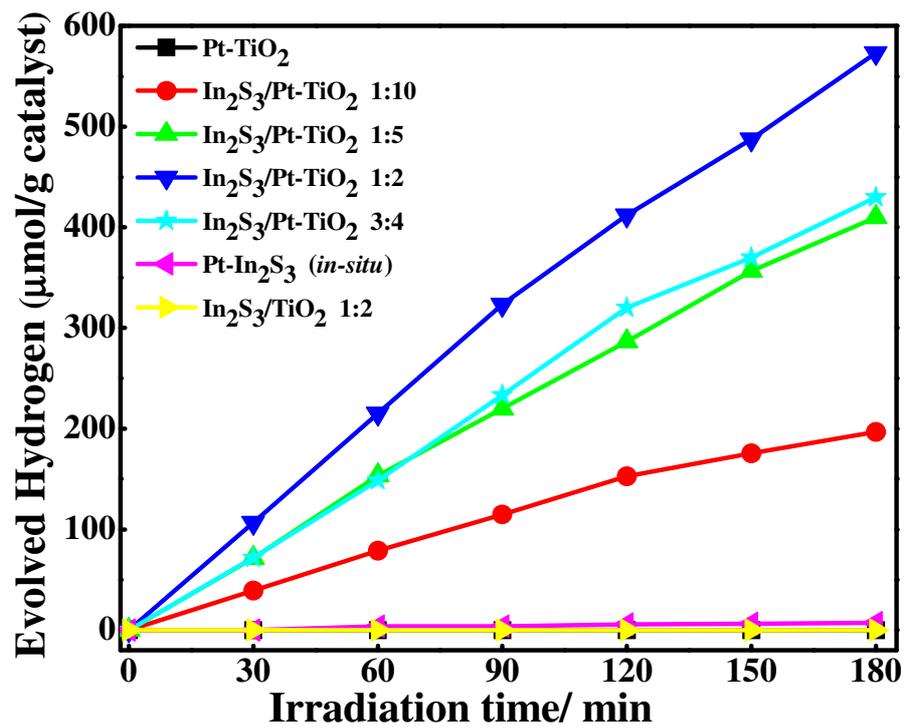

Fig. 4.

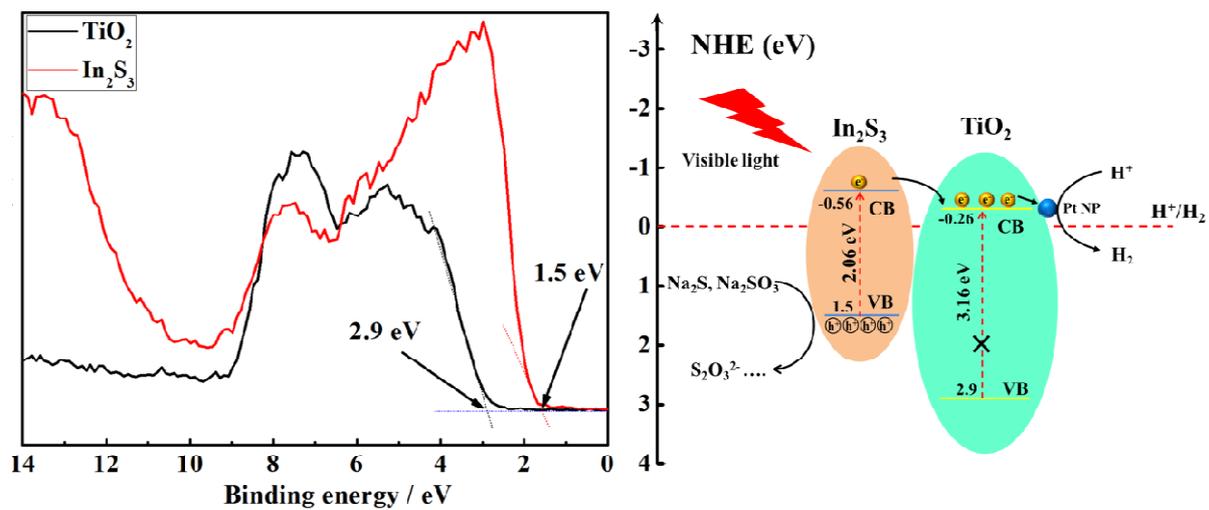

Fig. 5.

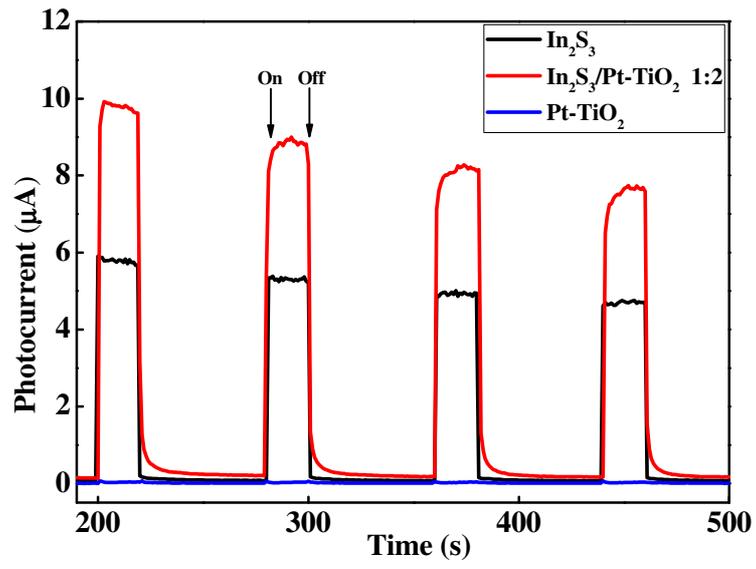

Fig. 6.

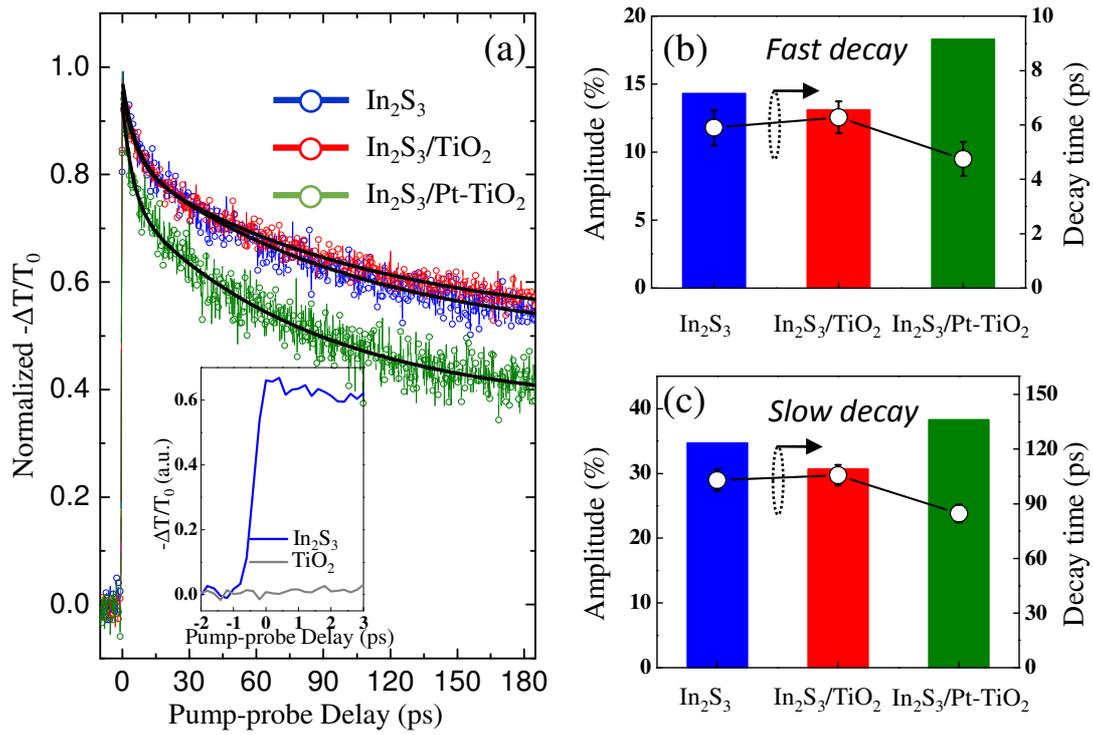

Fig. 7.

# Supporting Information

# Probing the Charge Separation Process on In$_2$S$_3$/Pt-TiO$_2$ Nanocomposites for Boosted Visible-light Photocatalytic Hydrogen Production

Fenglong Wang [a, b], Zuanming Jin [b, c], Yijiao Jiang [d,*], Ellen H. G. Backus [b], Mischa Bonn [b], Shi Nee Lou [a], Dmitry Turchinovich [b] and Rose Amal[a, *]

[a] School of Chemical Engineering, UNSW Australia, Sydney, NSW 2052, Australia.

[b] Max Planck Institute for Polymer Research, Ackermannweg 10, 55128 Mainz, Germany.

[c] Department of Physics, Shanghai University, 99 Shangda Road, Shanghai 200444, China.

[d] Department of Engineering, Macquarie University, Sydney, NSW 2109, Australia.

AUTHOR INFORMATION

**Corresponding Author**

Tel: +612-9385-4361, Email: r.amal@unsw.edu.au; yijiao.jiang@mq.edu.au

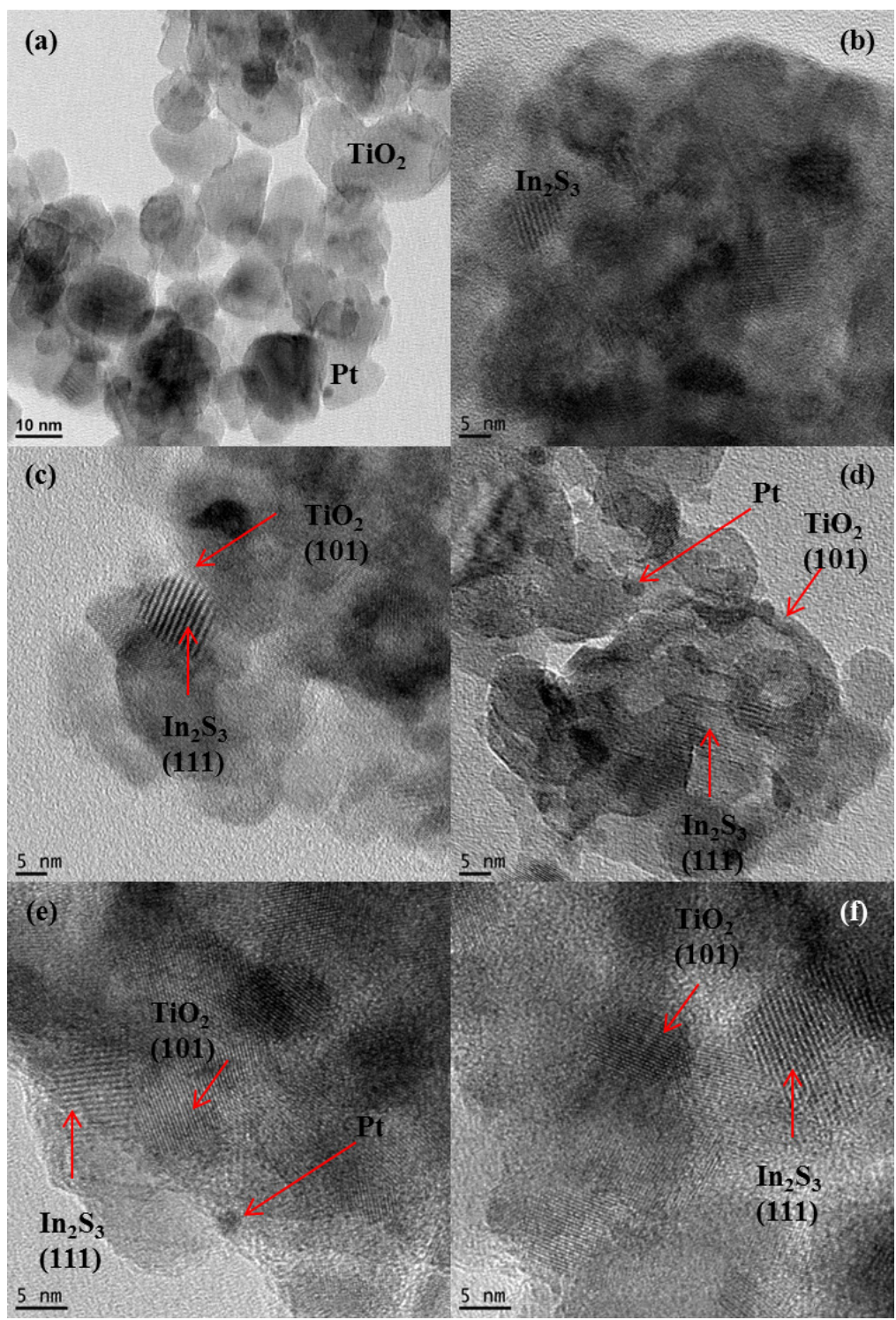

**Fig. S1** TEM images of (a) Pt-TiO$_2$, (b) In$_2$S$_3$, (c) In$_2$S$_3$/TiO$_2$, (d) In$_2$S$_3$/Pt-TiO$_2$ 1:10, (e) In$_2$S$_3$/Pt-TiO$_2$ 1:5 and (f) In$_2$S$_3$/Pt-TiO$_2$ 3:4.

Fig. S1 (a) clearly shows that spherical-shaped Pt nanoparticles with size around 2~3 nm are well distributed on the surface of $TiO_2$ support. Fig. S1 (b) displays the HRTEM image of $In_2S_3$ exhibiting the characteristic (111) facet with lattice spacing of 0.6 nm. Fig. S1 (c) shows the nanocomposites between $In_2S_3$ and $TiO_2$, in which the heterojunction between the two components are identified. Fig. S1 (d), (e) and (f) demonstrate the nanocomposites between $In_2S_3$ and Pt-$TiO_2$ with different ratios. By the characteristic spacing values of $In_2S_3$ (111) (0.6 nm) and $TiO_2$ (101) (0.35 nm) facets, it is easy to distinguish the phase-junctions in these composite materials. The TEM images confirm that the hetero-phased composites were successfully prepared.

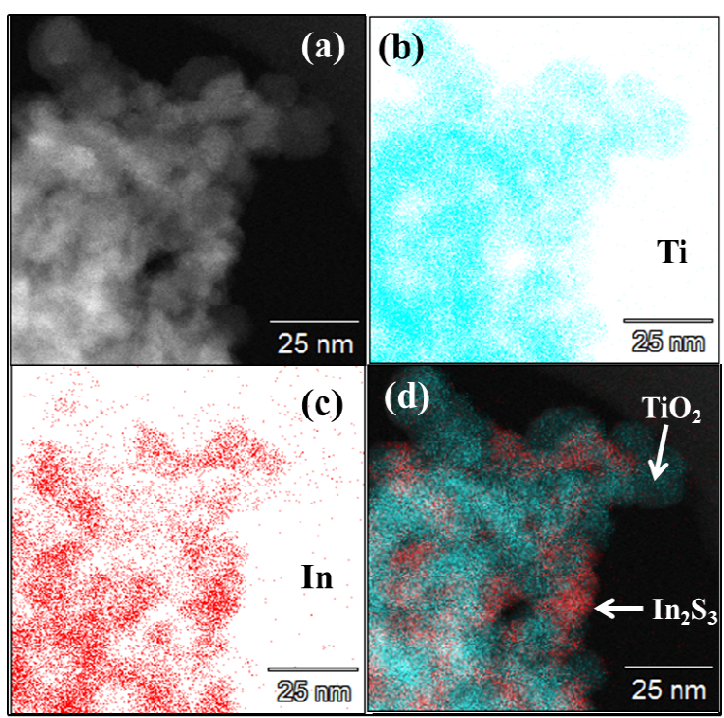

**Fig. S2** (a) TEM of $In_2S_3$/Pt-$TiO_2$ with mass ratio of 1:2, (b) Ti EDX mapping, (c) In EDX mapping, and (d) Ti EDX overlapping with In EDX mapping on the TEM image showing the phase interfaces between $TiO_2$ and $In_2S_3$.

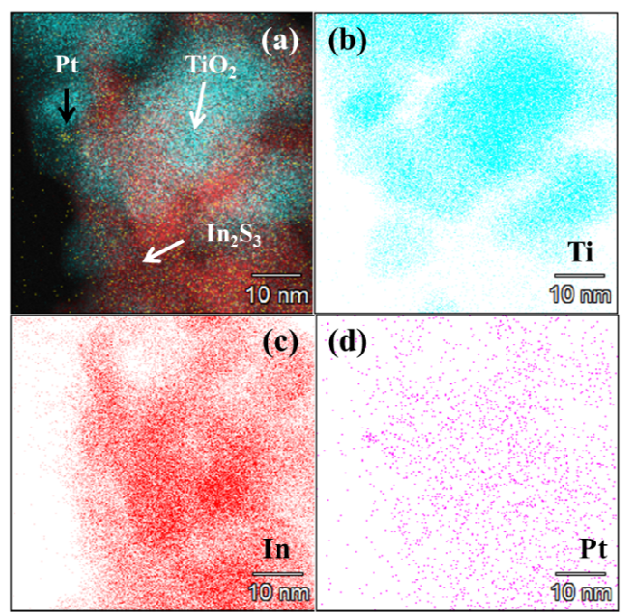

**Fig. S3** (a) Ti EDX, In EDX and Pt EDX mapping on the TEM image showing the phase interfaces in the composites, (b) Ti EDX mapping, (c) In EDX mapping, and (d) Pt EDX mapping.

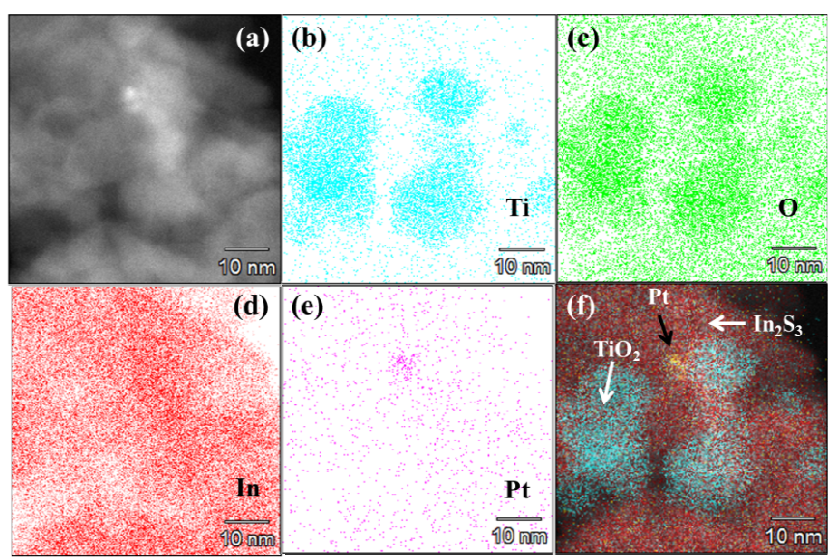

**Fig. S4** (a) STEM of $In_2S_3$/Pt-$TiO_2$ with mass ratio of 1:2, (b) Ti EDX mapping, (c) O EDX mapping, (d) In EDX mapping, (e) Pt EDX mapping and (f) Ti EDX, In EDX and Pt EDX mapping on the TEM image showing elemental distributions in the composites.

To show how $In_2S_3$ nanoparticles distributed on Pt-$TiO_2$ particle surfaces, we checked our sample under STEM and identified the phased by EDX elemental mapping, as shown in the following Fig. S2-4. These figures clearly display the phase distributions in $In_2S_3/Pt-TiO_2$ with mass ratio of 1:2 and one can conclude that hybridized nanocomposites were formed.

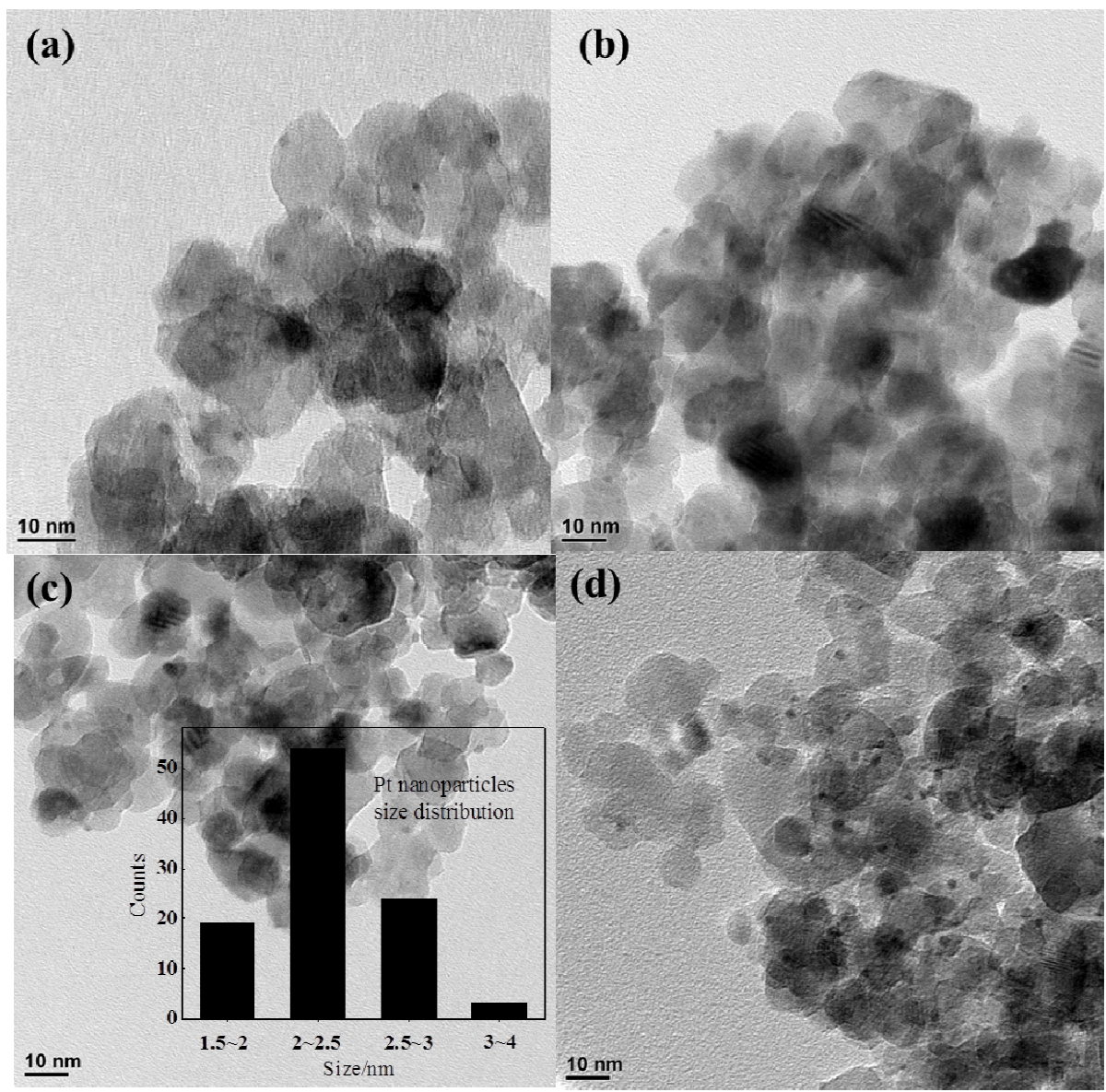

**Fig. S5** TEM images of Pt-TiO$_2$ nanocomposites showing the Pt nanoparticles with size around 2~3 nm are well distributed on TiO$_2$ surface. The inset of (c) shows the size distribution of Pt nanoparticles based on 100 particles measurement.

**Photocurrent response**

To better evaluate the photo-induced charge separation, the measured photocurrents are presented in terms of per mg of the active component ($In_2S_3$), as shown in Fig. S1. Under visible light irradiation, the photocurrent generated by the $In_2S_3$/Pt-$TiO_2$ was about five times higher than the bare $In_2S_3$. The finding reveals that the photo-excited electrons in the $In_2S_3$ conduction band can smoothly transfer to the $TiO_2$ conduction band, leading to enhanced charge separation. Compared to the bare $In_2S_3$, the higher photocurrent for the $In_2S_3$/$TiO_2$ demonstrates that the photo-generated electrons can easily migrate from $In_2S_3$ to $TiO_2$, indicating that the band alignment between these two semiconductors favours charge separation.

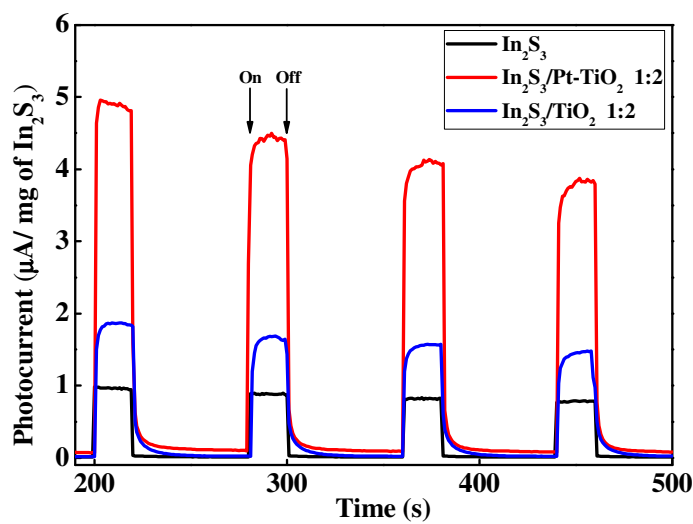

**Fig. S6.** Photocurrent responses of the $In_2S_3$, $In_2S_3$/$TiO_2$ and $In_2S_3$/Pt-$TiO_2$ nanocomposites under visible light irradiation ($\lambda > 420$ nm).

**XPS core-level spectra**

The XPS core-level spectra were recorded to investigate the valence state and the chemical composition of the catalyst surface. As shown in Fig. S2a, the peaks at 445.4 eV and 445.1 eV can be attributed to the In $3d_{5/2}$ in $In_2S_3$ and $In_2S_3$/Pt-TiO$_2$, respectively, while the two peaks in the lower binding energy region arise from In $3d_{3/2}$. In the S core spectra shown in Fig. S2b, the peaks at 162.0 eV and 161.8 eV are related to the S $2p_{3/2}$ in $In_2S_3$ and $In_2S_3$/Pt-TiO$_2$, respectively.[1] Separations between the In and S spin orbitals are around 7.3 eV and 1 eV, and the ratios between the peak areas are found to be ca. 2:3 and 1:2, respectively, indicating that the In and S exist as $In^{3+}$ and $S^{2-}$ oxidation states in the nanocomposites.[2]

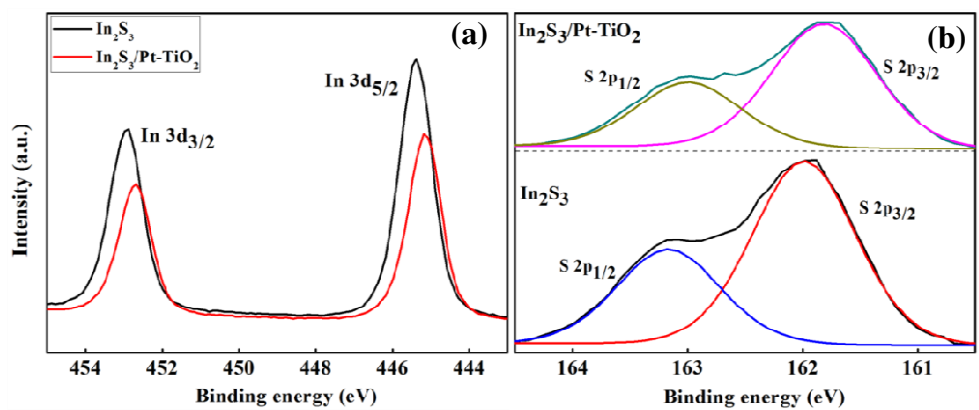

**Figure S7.** (a) In *3d* and (b) S *2p* core-level XPS spectra of $In_2S_3$ and $In_2S_3$/Pt-TiO$_2$.

**Activity comparison under both full spectrum and visible light illumination**

**Table S1.** The hydrogen generation properties for Pt-$In_2S_3$ and $In_2S_3$/Pt-TiO$_2$ under visible light and full spectrum irradiation.

| Light<br><br>Catalyst | Full spectrum irradiation | Visible light irradiation<br>(λ>420 nm) |
|---|---|---|
| Pt-$In_2S_3$ *(in-situ)* | 3.3 µmol | 2.1 µmol |
| $In_2S_3$/Pt-TiO$_2$ | 271 µmol | 172 µmol |

**Bi-exponential fittings for normalized real photoconductivity dynamics of the materials**

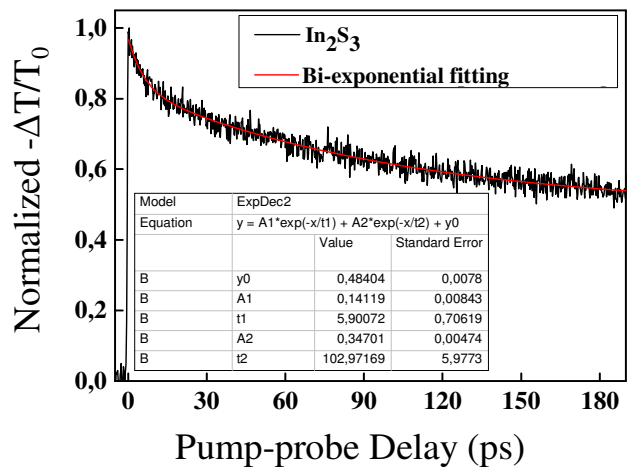

**Figure S8.** Bi-exponential fitting for normalized real photoconductivity dynamics of In$_2$S$_3$.

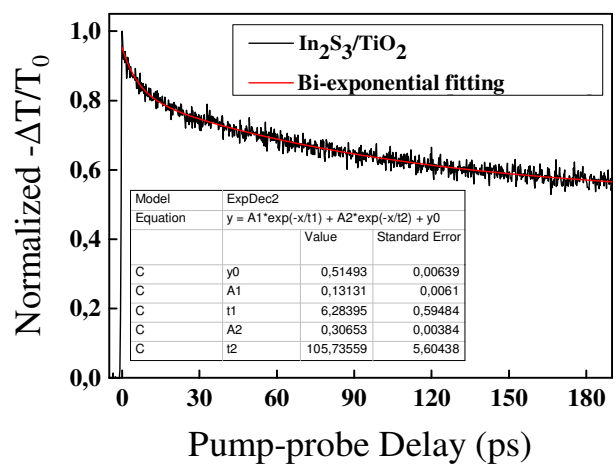

**Figure S9.** Bi-exponential fitting for normalized real photoconductivity dynamics of In$_2$S$_3$/TiO$_2$.

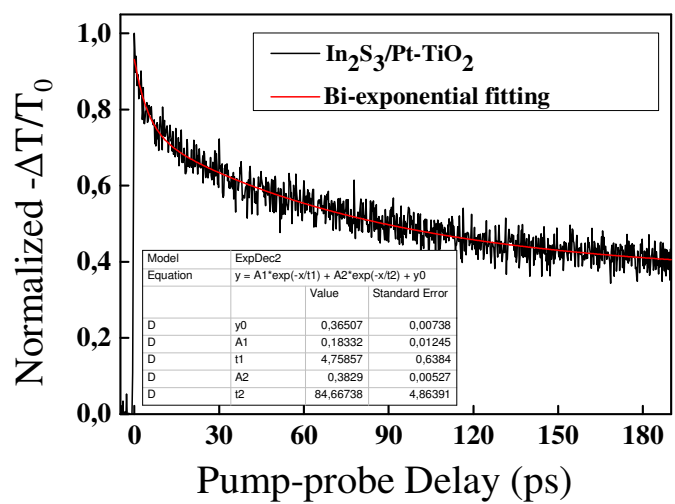

**Figure S10.** Bi-exponential fitting for normalized real photoconductivity dynamics of In$_2$S$_3$/Pt-TiO$_2$.